\documentclass[letterpaper,twocolumn,10pt]{article}

\usepackage{usenix}
\usepackage{graphicx}             
\usepackage{subfig}            
\usepackage{url}                  
\usepackage[T1]{fontenc}
\usepackage{nicefrac}             
\usepackage{indentfirst}          
\usepackage[super, negative]{nth}
\usepackage{amsfonts}
\usepackage{amssymb}
\usepackage{multirow}
\usepackage{balance}
\usepackage{xspace}
\usepackage{array,booktabs}
\usepackage{latexsym}
\usepackage{pifont}
\usepackage{colortbl}
\usepackage{bytefield}
\usepackage{listings}
\usepackage{algorithm}
\usepackage[noend]{algpseudocode}
\usepackage{xcolor}

\algrenewcommand\algorithmicindent{1.0em}%

\definecolor{my-gray}{gray}{0.45}
\algnewcommand{\ccomment}[1]{\State \textit{/* #1 */}}
\algnewcommand{\icomment}[1]{\Comment{\textit{\color{gray}#1}}}
\algnewcommand{\iccomment}[1]{\textit{\color{my-gray}{~/* #1} */}}

\algnewcommand\Or{~\textbf{or}~}
\algnewcommand\And{~\textbf{and}~}
\algnewcommand\In{~\textbf{in}~}
\algnewcommand\algorithmicswitch{\textbf{switch}}
\algnewcommand\algorithmiccase{\textbf{case}}
\algnewcommand\algorithmicassert{\texttt{assert}}
\algnewcommand\Assert[1]{\State \algorithmicassert(#1)}%
\algdef{SE}[SWITCH]{Switch}{EndSwitch}[1]{\algorithmicswitch\ #1\ \algorithmicdo}{\algorithmicend\ \algorithmicswitch}%
\algdef{SE}[CASE]{Case}{EndCase}[1]{\algorithmiccase\ #1}{\algorithmicend\ \algorithmiccase}%
\algdef{SE}[DEFAULTCASE]{DefaultCase}{EndDefaultCase}{\textbf{default}}{\ }
\algtext*{EndSwitch}%
\algtext*{EndCase}%
\algtext*{EndDefaultCase}%

\newcommand{\dfn}[1]{\textit{#1}}            

\newcommand{\mmb}[0]{\texttt{mmb}\xspace}

\definecolor{mygreen}{rgb}{0,0.6,0}

\lstdefinestyle{BashInputStyle}{
  language=bash,
  basicstyle=\small\sffamily,
  frame=tb,
  columns=fullflexible,
  linewidth=1\linewidth,
  breaklines=true,
}

\begin{document}

\title{mmb: Flexible High-Speed Userspace Middleboxes}

\author{
{\rm Korian Edeline, Justin Iurman, Cyril Soldani, Benoit Donnet}\\
Universit\'e de Li\`ege\\
Montefiore Institute\\
Li\`ege, Belgium
}

\maketitle

\subsection*{Abstract}
Nowadays, Internet actors have to deal with a strong increase in Internet
traffic at many levels.  One of their main challenge is building high-speed and
efficient networking solutions.  In such a context, kernel-bypass I/O frameworks
have become their preferred answer to the increasing bandwidth demands.  Many
works have been achieved, so far, all of them claiming to have succeeded in
reaching line-rate for traffic forwarding.  However, this claim does not hold
for more complex packet processing. In addition, all those solutions share
common drawbacks on either deployment flexibility or configurability and
user-friendliness.

This is exactly what we tackle in this paper by introducing \mmb, a VPP
middlebox plugin.  \mmb allows, through an intuitive command-line interface, to
easily build stateless and stateful classification and rewriting middleboxes.
\mmb makes a careful use of instruction caching and memory prefetching, in addition to other techniques used by other high-performance I/O frameworks. We compare \mmb performance with other performance-enhancing middlebox solutions, such as kernel-bypass framework, kernel-level optimized approach and other state-of-the-art solutions for enforcing middleboxes policies (firewall, NAT, transport-level engineering). We demonstrate that \mmb performs, generally, better than existing solutions, sustaining a line-rate processing while performing large numbers of complex policies.

\section{Introduction}\label{intro}
Global Internet traffic has constantly increased over the past decade.
In 2017, 17.4 billions of devices have generated more than 45,000 GB/second
Internet traffic. By 2022, the number of devices connected to IP networks will
reach 28.5 billions, among which 51\% will follow a machine to machine (M2M)
communication scheme, and their traffic will attain 150,700 GB/second, with hours
peaking up to a x4.8 increase factor. Following the introduction of
of 4K, Ultra-High-Definition (UHD), or video streaming, Video traffic, which is
particularly delay-sensitive, will account for 82 \% of this
total~\cite{gerber2011traffic,cisco2018cisco}. The fixed broadband and the mobile
network speed will also continue to grow, to an average of respectively 75.4 Mbps
and 28.5 Mbps.

In parallel, the traditional TCP/IP architecture (i.e., the end-to-end
principle) is becoming outdated in a wide range of network situations.  Indeed,
corporate networks~\cite{sherry2012making}, WiFi hotspots, cellular
networks~\cite{wang2011untold}, but also Tier-1 ASes~\cite{edeline2017first} are
deploying more and more \dfn{middleboxes} in addition to traditional network
hardware.  A middlebox is a network device inspecting, filtering, or even
modifying packets that traverse it.  Typically, a middlebox performs actions on
a packet that are different from standard functions of an IP router.  Indeed,
middleboxes may be deployed for, e.g.,  security (IDS, NATs, firewalls)
and network performance (load balancer, WAN optimizer).

Internet actors have thus to deal with this double increase at many levels, and
particularly, in building high-speed networking solutions. A wide range of
Kernel-bypass I/O frameworks are available to answer this increasing bandwidth
demand, and the Linux kernel has been striving to stay
afloat~\cite{hoiland2018express,brouer2018xdp}. Many of
those efforts claims to have succeeded in reaching line-rate for traffic
forwarding, less so for more complex packet processing (e.g., a firewall with a
large number of rules, TCP options). Moreover, all of them share common
drawbacks, on either deployment flexibility by necessitating expensive hardware
or specific OS to maintain reasonable performances, or configurability and
user-friendliness by requiring non-trivial programming for basic adaptation of
common network functions.

In this paper, we want to overcome those limitations by introducing \mmb
(\textbf{\underline{M}}odular
\textbf{\underline{M}}iddle\textbf{\underline{B}}ox), an open-source~\footnote{Due to double-blind submission, a link to
the source code will be provided if the paper is accepted.} extension to the
Vector Packet Processing (VPP~\cite{linguaglossa2017high}) high-speed
kernel-bypassing framework.  \mmb aims at achieving line-rate forwarding
performance while performing a large number of complex packet manipulation
defined by middlebox policies. It leverages VPP employment of classical and
recent advances in packet processing techniques, such as computation and I/O
batching, Zero-Copy forwarding, low-level parallelism, and caching efficiency.

Moreover, by implementing combinable generic middlebox policies, configurable
from an intuitive command-line interface, \mmb allows for out-of-the-box
middlebox deployment and easy adaptation. On modern hardware, it is able to hold
baremetal-like performance while running on a virtual machine, thanks to PCIe
passthrough technologies (i.e., SR-IOV, virtio). Finally, \mmb benefits from VPP
continuous development and maintenance.

We conduct a thorough comparison of trending high-performance packet processing
solutions with \mmb, for a selection of simple to complex use cases, packet
forwarding, firewall-like packet filtering, packet filtering with stateful
flow tracking, packet filtering with stateful matching and packet mangling
(i.e., NAT), and TCP options filtering and mangling. We find that, with few
hardware restrictions and without the need to write a single line of code,
\mmb is able to sustain packet forwarding at line-rate speed when enforcing
a large set of diverse and complex classification and mangling rules, while
other solutions either perform worse, require specific hardware or OS,
necessitate expert-level configuration, or have inner design limitations that
make them inapplicable.

The remainder of this paper is organized as follows: Sec.~\ref{vpp} provides the
required background for this paper, focusing on the VPP framework and its main
features employed by \mmb;  Sec.~\ref{architecture} introduces \mmb by detailing
its architecture and important aspects; Sec.~\ref{performance} contains the
performance evaluation of \mmb. It describes and motivates the selected use
cases, introduces the state-of-the-art tools compared to \mmb, explains their
configurations, and discusses the results; Sec.~\ref{related} positions \mmb
with respect to notable high-performance packet processors; finally,
Sec.~\ref{conclusion} concludes this paper by summarizing its main achievements.

\section{VPP Background}\label{vpp}
This section aims at providing the required background for \mmb.  In particular,
it focuses on \dfn{Vector Packet Processing} (VPP)~\cite{vpp}, a 17-year old
Cisco-developed technology providing a high-performance, extensible,
feature-rich, packet-processing stack that runs in user space. It implements
a full network stack and is designed to be customizable. It can run on I/O frameworks
such as DPDK~\cite{dpdk-link}, Netmap~\cite{rizzo2012netmap}, or OpenDataPlane
(ODP)~\cite{odp}.

VPP leverages techniques such as batch-processing, Receive-side Scaling (RSS) queues, Zero-Copy by allowing userspace applications to have Direct Memory Access (DMA) to the memory region used by the NIC, offloading certain packet processing functions to dedicated hardware, and I/O batching to reduce the overhead of NIC-initiated interrupts. While those techniques have been implemented in other kernel-bypassing frameworks (e.g., FastClick~\cite{barbette2015fast}) and have been shown to drastically improve performances~\cite{barbette2018building}, VPP attempts to surpass it by introducing parallel processing on multiple CPU cores, to maximise hardware instruction pipelining, alongside an optimal use of CPU caches to minimize the memory access bottleneck. To this end, VPP introduces particular coding practices (e.g.: memory prefetching, cache-fitting processing nodes, branch
prediction) to maximize low-level parallelism and cache locality.

\begin{figure}[!t]
   \begin{lstlisting}[language=C,basicstyle=\footnotesize,frame=single,commentstyle=\color{mygreen}]
   while(n_left_from >= 2){
      /* prefetch next iteration */
      if(PREDICT_TRUE(n_left_from >= 4)){
         vlib_prefetch_buffer_header(b[2], STORE);
         vlib_prefetch_buffer_header(b[3], STORE);
      }

      process(b[0]);
      process(b[1]);

      b += 2;
      next += 2;
      n_left_from -= 2;
   }

   /* process remaining packets */
   while(n_left_from > 0){
      process(b[0]);

      b += 1;
      next += 1;
      n_left_from -= 1;
   }
   \end{lstlisting}
   \caption{VPP Dual-loop.}
   \label{vpp:optimization.loop}
\end{figure}

\subsection{Processing Nodes}\label{vpp:nodes}
The VPP packet processing path is based on a directed \emph{forwarding graph}
architecture. An example of such graph is shown in Fig.~\ref{fig.nodes}, that
illustrates the subgraph used by \mmb. It is made of small, modular \emph{nodes}
performing a set of functions (e.g., \texttt{dpdk-input}, \texttt{ip4-lookup})
to packets, in userspace.  Each node is designed to entirely fit inside the instruction cache. The graph node dispatcher is responsible for directing the
packet vectors through the graph. This modular architecture enables the
development of independent \emph{plugins}, that consist in shared libraries
loaded at runtime, to plug their own nodes into the existing VPP forwarding
graph, or rearrange it.

\subsection{Packet Vectors}\label{vpp:vector}
Kernel-bypass frameworks usually rely on a classical run-to-completion
approach~\cite{barbette2015fast,kim2012power}, where each single packet is
processed by each function separately. However, this approach has a few
downsides related to memory access, which is a well-known bottleneck of software
performance~\cite{boncz1999database}, ultimately resulting in reduced
throughput.

First and foremost, the instruction cache hit rate suffers from continually having  to load the different packet processing functions, and even more so when the cache is full. Moreover, when processing a packet, there is no a-priori information on the next function to be executed, neither which portions the packet data will be read. This prevents the establishment of prefetching strategies, which are efficient in addressing the memory access bottleneck.

VPP chooses to rely on a per-node batch processing, by systematically using
pre-allocated packet batches (i.e., \emph{vectors}), that contains typically up
to 256 packets. When a function is applied to a packets batch, the first packet
causes the function to be loaded in the instruction cache. Then, the following
packets are guaranteed to hit the cache, amortizing the cost of the initial
cache miss over the whole packet vector. Moreover, this approach gives a-priori
information on the next data sections to be read, allowing for efficient
prefetching strategies.

\subsection{Low-level Optimization}\label{vpp:optimization}
VPP heavily encourages developers to set up multiple coding practice in the form
of low-level optimizations techniques, in order to exploit all available hardware
optimizations, and ultimately to improve the framework throughput.

Most VPP packet processing functions are \emph{N-loops}, which consists in
explicit handling $N$ packets per iteration to increase the code parallelism.
This practice aims at exploiting CPU hardware pipelining and at amortizing the
cost of instruction cache misses. Fig.~\ref{vpp:optimization.loop} shows an
example of a dual-loop packet processing algorithm with data prefetching. By
unrolling the loop, this algorithm allows for subtracting the processing time
of two packets to the fetching time of the two next packets buffers. VPP authors
notes that quad-loops (i.e., $N$=4) are only beneficial for small processing
functions, because the introduced space-time trade-off is bounded by the limited
size of instruction L1 caches~\cite{linguaglossa2017high}.

VPP also relies on \emph{branch prediction} by having programmers to give
hints on the probability of code branches. This practice benefits, in case of a
correct branch prediction, of avoiding a pipeline reset. Another important
practice is allocating $N$ buffers at a time rather than individually, and when
possible, pre-allocating them.

\section{Modular Middlebox}\label{architecture}

\begin{figure}[!t]
  \begin{center}
    \hspace*{-0.5cm}\includegraphics[width=9.5cm]{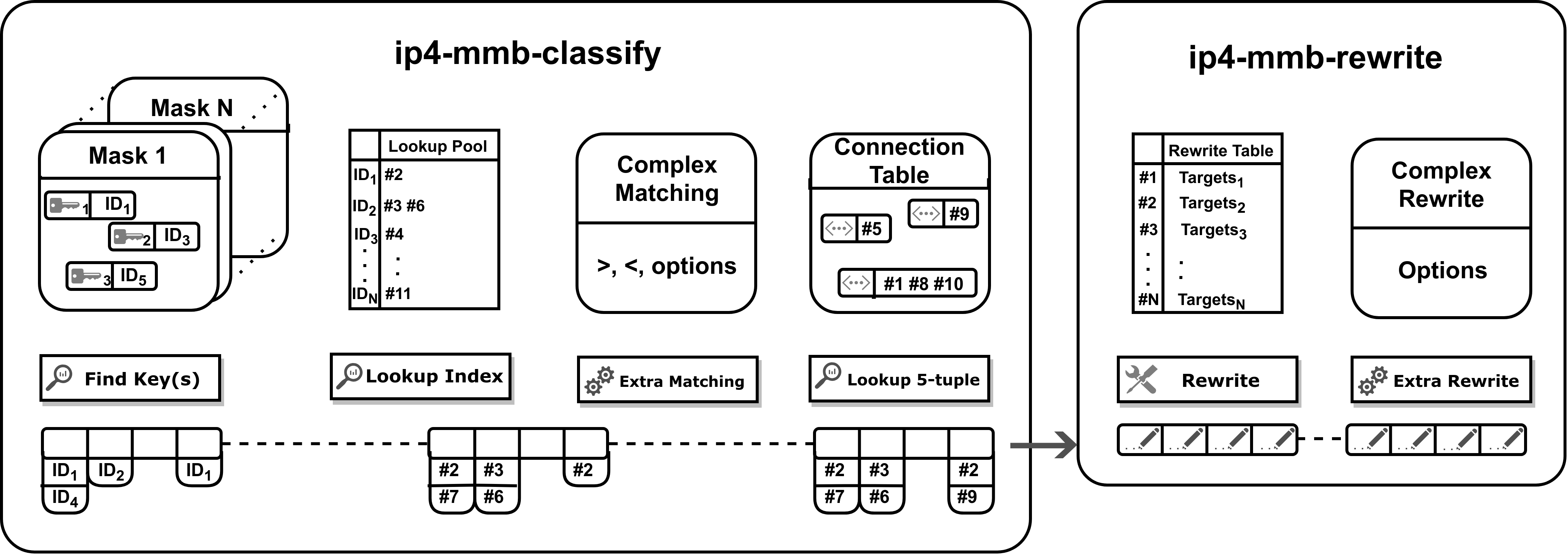}
  \end{center}
  \caption{\mmb processing path.}
  \label{fig.processing}
\end{figure}

\mmb (\textbf{\underline{M}}odular
\textbf{\underline{M}}iddle\textbf{\underline{B}}ox) is a VPP extension that
performs stateless and stateful classification, and rewriting. It achieves
stateless packet matching based on any combination of constraints on network or
transport protocol fields, stateful TCP and UDP flow matching, packet mangling,
packet dropping and bidirectional mapping. \mmb is partly protocol-agnostic by
allowing to match and rewrite fields \texttt{ip4-payload}, \texttt{udp-payload},
and \texttt{tcp-opt}, and allows for on-the-fly configuration.

Sec.~\ref{architecture.overview} provides a global overview of \mmb, while
Sec.~\ref{architecture:classification} and Sec.~\ref{architecture:rewrite} focus
on \mmb nodes for classification and rewriting operations.

\subsection{General Overview}\label{architecture.overview}
\begin{figure}[!t]
  \begin{center}
    \includegraphics[width=7.5cm]{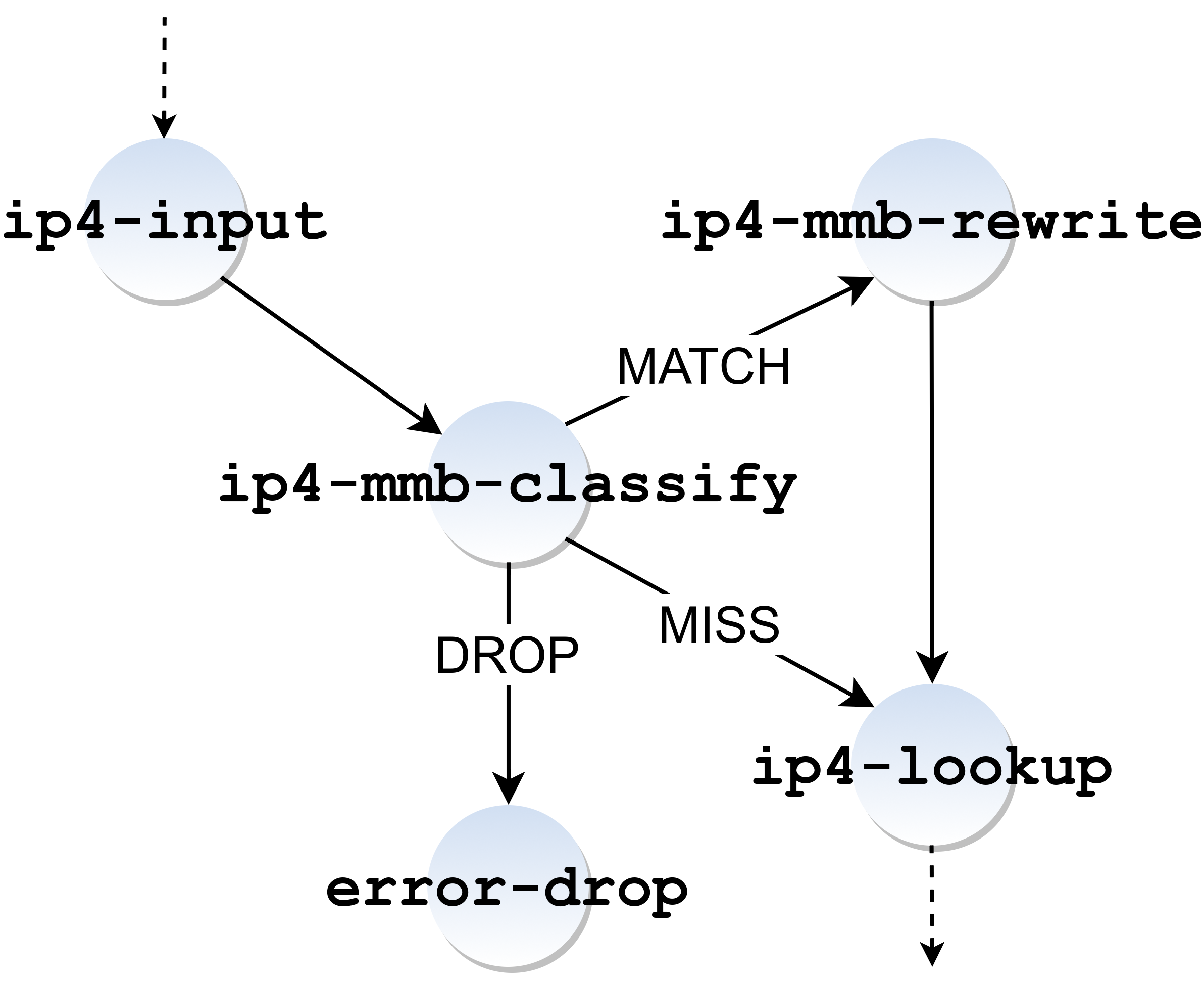}
  \end{center}
\vspace*{-0.5cm}
  \caption{VPP forwarding graph with \mmb nodes.}
  \label{fig.nodes}
\end{figure}

Following the VPP architecture (see Sec.~\ref{vpp}), \mmb forwarding graph
consists in two nodes, a \emph{classification}
(e.g., \texttt{ip4-mmb-classify}) and a \emph{rewrite} node (e.g.,
\texttt{ip4-mmb-rewrite}), as shown in Fig.~\ref{fig.nodes}. When \mmb
is enabled, its nodes are connected to the processing graph. If IPv6 is
available, the IPv6 variants of those two nodes can also be connected to
the IPv6 forwarding graph. The classification node is placed right after
the \texttt{ip4-input} (or \texttt{ip6-input}) node, that validates the
IP4 header checksum, verifies its length and discards packets with expired TTLs.

Depending on the outcome of the classification step, that can either be
\emph{drop}, \emph{miss}, or \emph{match}, packets are forwarded
respectively to the \texttt{error-drop} node which will discard them,
to \texttt{ip4-lookup}, the node responsible for the Forwarding Information Base  (FIB) lookups, that then dispatches packets to the corresponding processing path, or the \texttt{mmb-rewrite} node, applying modification rules to packets.

Overall, \mmb consists in three processing paths that can each be traversed or
not by packet vectors, depending on the input policies. A fast path, which
relies on VPP bounded index hash tables and implements the mask-based matching
operation using binary operators, is shown in Fig.~\ref{fig.formula}. This path
is enabled when a rule without any TCP option is entered. Moreover, it restricts
the conditions to \textbf{==} (isequal). The stateful flow matching is using this fast path. A first slow path for rules that uses complex conditions ($\ne$, $<$,  $>$, $\leq$, $\geq$), and a second slow path for the linked list parsing required  when classifying based on TCP options as well as when rewriting them.

\begin{figure}[!t]
  \begin{center}
    \hspace*{-0.5cm}\includegraphics[width=9.5cm]{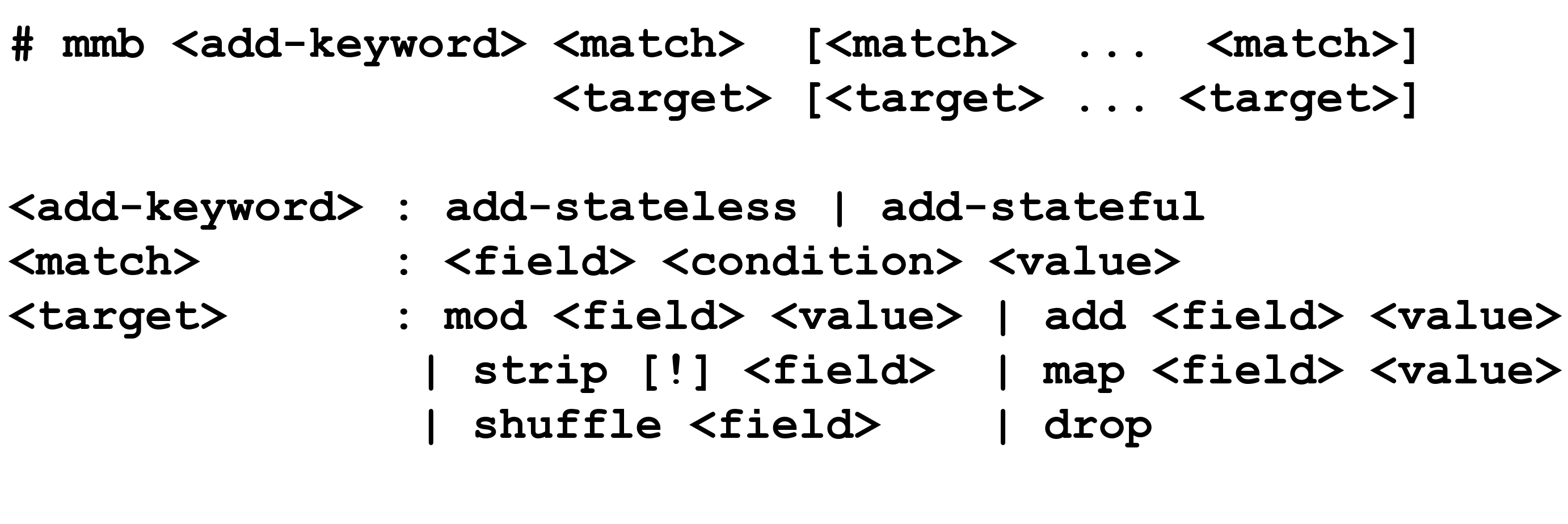}
  \end{center}
\vspace*{-0.5cm}
  \caption{\mmb command-line interface syntax.}
  \label{fig.syntax}
\end{figure}

The main goal of \mmb is to be easily configurable, and to allow defining
a wide range of middlebox policies by combining rules, defined by using commands with
a generic semantic~\cite{edeline2015towards,edeline2017observation}, at high-speed.
To this end, we define a grammar (see Fig.~\ref{fig.syntax}) that can be used to
build a packet processing middlebox directly from a command-line interface.
For example, building a middlebox that rewrites TCP port 80 to port 443 is done
as follows:
\begin{lstlisting}[style=BashInputStyle]
vpp# mmb add tcp-dport 80 mod tcp-dport 443
\end{lstlisting}
Here is another example of a middlebox stripping all options but MSS and WSCALE if the packet
contains the timestamp option:
\begin{lstlisting}[style=BashInputStyle]
vpp# mmb add tcp-opt-timestamp strip ! tcp-opt-mss strip ! tcp-opt-wscale
\end{lstlisting}
A full usage of the command-line interface with examples is provided with \mmb
source code.\footnotemark[1]

\subsection{Classification Node}\label{architecture:classification}
\mmb packet processing is displayed in Fig.~\ref{fig.processing}. The
classification node is an extension to VPP classification module, that consists
in four distinct steps: a mask-based constraint matching step, an index lookup
pool, a complex matching step and a connection table.

The mask-based matching determines if each packet satisfies constraints
on fixed offset fields. For this, we create one classification table per packet
mask (e.g., per combination of fields in the match constraint), sized from 16
bytes to at most 80 consecutive bytes. We create one key for a given table per
value for its associated packet mask. For each table, the search for a key
matching a given packet is a hash-based search performed in constant time.

\begin{algorithm}[!t]
    \caption{Matching operation}
    \label{alg:match}
    \begin{algorithmic}
      \Function{Match}{$pkt$, $mask$, $key$, $skip$, $chunks$}
          \State $res \gets (pkt[skip]\ \&\ mask[\texttt{0}]) \oplus key[\texttt{0}]$

           \Switch{$chunks$}
             \Case{$\texttt{5}$}
               \State $res \gets res\ |\ ((pkt[skip+\texttt{4}]\ \&\ mask[\texttt{4}]) \oplus key[\texttt{4}])$
             \EndCase
             \Case{$\texttt{4}$}
               \State $res \gets res\ |\ ((pkt[skip+\texttt{3}]\ \&\ mask[\texttt{3}]) \oplus key[\texttt{3}])$
             \EndCase
             \Case{$\texttt{3}$}
               \State $res \gets res\ |\ ((pkt[skip+\texttt{2}]\ \&\ mask[\texttt{2}]) \oplus key[\texttt{2}])$
             \EndCase
             \Case{$\texttt{2}$}
               \State $res \gets res\ |\ ((pkt[skip+\texttt{1}]\ \&\ mask[\texttt{1}]) \oplus key[\texttt{1}])$
             \EndCase
             \Case{$\texttt{1}$}
               \State $break$
             \EndCase
             \DefaultCase
               \State $abort()$
             \EndDefaultCase
           \EndSwitch

          \If{$zero\_byte\_mask(res) = \texttt{0xffff}$}
            \State \Return $\texttt{1}$
          \Else
              \State \Return $\texttt{0}$
          \EndIf
      \EndFunction
    \end{algorithmic}
\end{algorithm}

The matching operation consists of two binary operations (AND and XOR), as shown
in Fig.~\ref{fig.formula}.\ref{fig.formula.1}, which are applied to
consecutive chunks of 16 bytes, starting from the first non-zero byte in the
mask. Each results are OR'ed into a 16-byte variable, that is compared to zero
to verify if the matching operation was successful. This operation is illustrated
in Alg.~\ref{alg:match}.

Then, for each packet that matched at least one mask-key combination, \mmb
checks if an additional matching is needed, with a constant-time lookup, and
performs it. Additional matching is necessary for constraints on linked-list
based fields such as TCP options.

Finally, each packet is matched to a connection table via its 5-tuple.
The connection tables keep track of every connection that matched at least
one stateful rule, and implements a flag tracking and a timeout mechanism
without interruptions. This allows, for example, for reflexive policies. Both
TCP and UDP are handled by the connection table.

If a packet matches at least one rule with a drop target, it is
immediately forwarded to the \texttt{error-drop} node. If the packet matches
only non-drop rules, it is forwarded to the \texttt{mmb-rewrite} node, and if the packet does not match any rule, it is handed to the next non-\mmb node,
\texttt{ip4-lookup}.

\subsection{Rewrite Node}\label{architecture:rewrite}
The \texttt{mmb-rewrite} node consists in two operations: a mask-based
rewrite step that works on the fixed offset fields, similarly to the first
step of the classification node, and a complex rewrite step for linked-list
based fields.

To perform the rewrite operation, or application of targets, we build a target
mask and a target key when the rule is added. The rewrite is then performed with  two binary operations (AND and OR), as shown in
Fig.~\ref{fig.formula}.\ref{fig.formula.2}.

\mmb allows to perform packet mangling by defining, for any rule, a set of
static and dynamic targets. Static targets consists in setting a user-defined
value to a chosen field. In the case of TCP options, targets may also define
an option strip or an addition. Dynamic targets allows for setting a different
value, within a predefined value range or random, on a per-connection basis.

\begin{figure}[!t]
   \begin{equation}\label{fig.formula.1}
     Result_{Classif.} = (Packet\ \&\ Mask)\ \oplus\ Key
   \end{equation}
   \begin{equation}\label{fig.formula.2}
     Result_{Rewrite} = (Packet\ \&\ Mask)\ |\ Key
   \end{equation}
\caption{Binary operations for packet classification and rewrite.}
\label{fig.formula}
\end{figure}

\section{Performance}\label{performance}
In this section, we evaluate \mmb performances. We describe and
motivate the selected use cases, introduce the state-of-the-art
kernel-bypass frameworks and other tools that we compare to and we explain
how we configured them. Finally, we discuss the obtained results.

\subsection{Testbed Description}\label{performance.testbed}
The testbed consists of three machines with Intel Xeon CPU E5-2620 v4 @ 2.10GHz,
8 Cores, 16 Threads, 32GB RAM, running Debian 9.0 with 4.9 kernels. Two of these
machines play the role of Traffic Generators (TGs), while one is the Device
Under Test (DUT). An additional machine with Intel Xeon CPU E5-2630 v3 @ 2.40GHz
8 Cores, 16 Threads, 16GB RAM, running Ubuntu Server 18.04.1 with 4.15 kernel,
is used as alternative DUT for experiments requiring a more recent kernel.
Each machine is equipped with a Intel XL710 2x40GB QSFP+ NIC with Receive-Side
Scaling enabled (RSS) connected to a Huawei CE6800 switch using one port each
for TGs and both for the DUT.

The DUT runs VPP 18.10, DPDK 18.08 with 10 1-GB huge pages, and a
kvm/QEMU 2.8.1 hypervisor with a Ubuntu 18.04 guest. The TGs run
iperf3~\cite{iperf}, nginx 1.10.3 and wrk 4.0.2.

The DUTs are configured to maximize their performances. The scaling
governor is set to run the CPU at the maximum frequency. 7 cores out of 8
(14 threads) are
isolated from the kernel scheduler to make sure that no other tasks is being run
on the same physical CPU, and pinned to the process under test. We enable
adaptive-ticks CPUs to omit unnecessary scheduling-clock ticks for CPUs with
only one runnable task, which we ensure by setting the CPU affinity for VPP,
and we enable Read-Copy-Update (RCU) callback offloading.

\begin{figure}
    \centering
    \subfloat[Direct]{
      \centering
      \includegraphics[width=4.0cm]{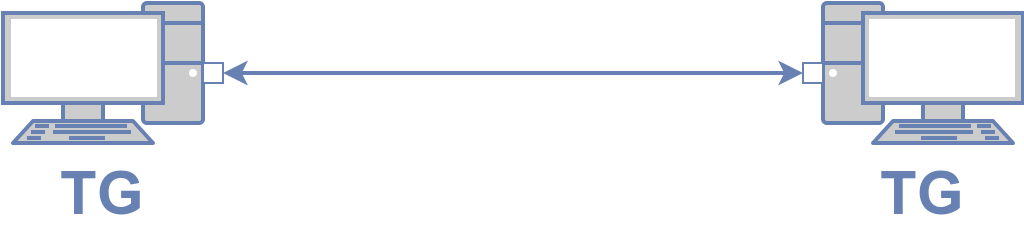}
      \label{fig.performance.setup.direct}
    }
    \subfloat[Indirect]{
      \centering
      \includegraphics[width=4.0cm]{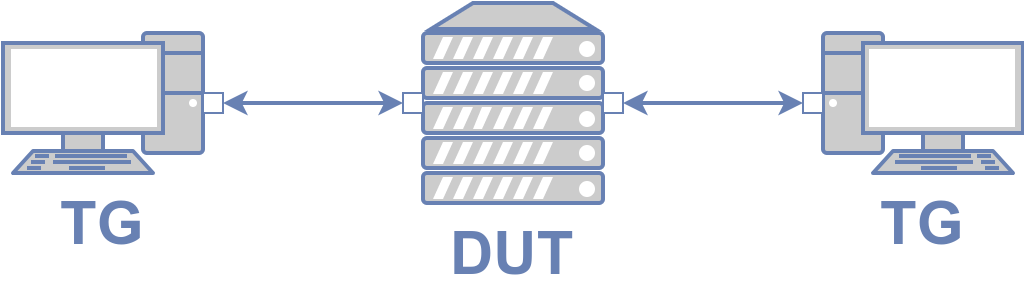}
      \label{fig.performance.setup.indirect}
    }

    \subfloat[PCI Passthrough]{
      \centering
      \includegraphics[width=4.0cm]{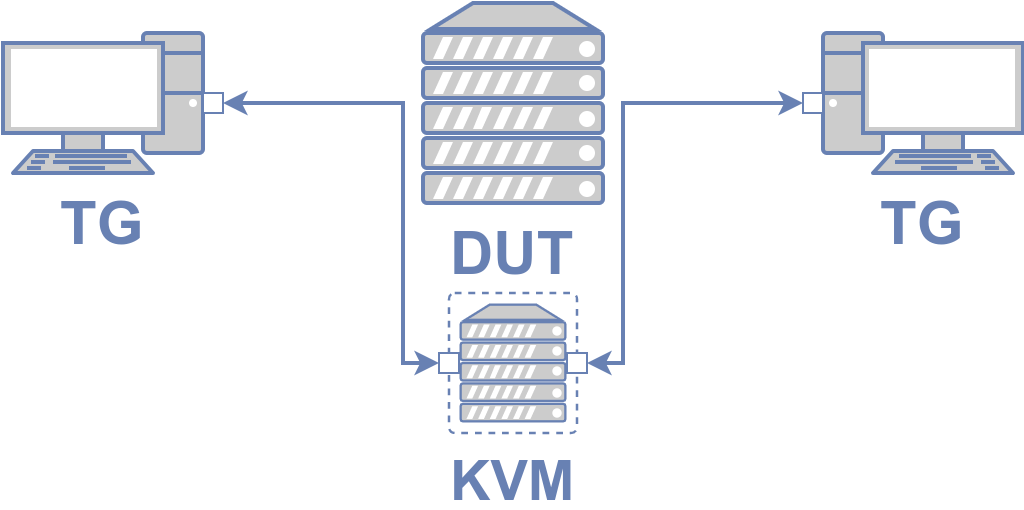}
      \label{fig.performance.setup.passthru}
    }
    \subfloat[Bridged]{
      \centering
      \includegraphics[width=4.0cm]{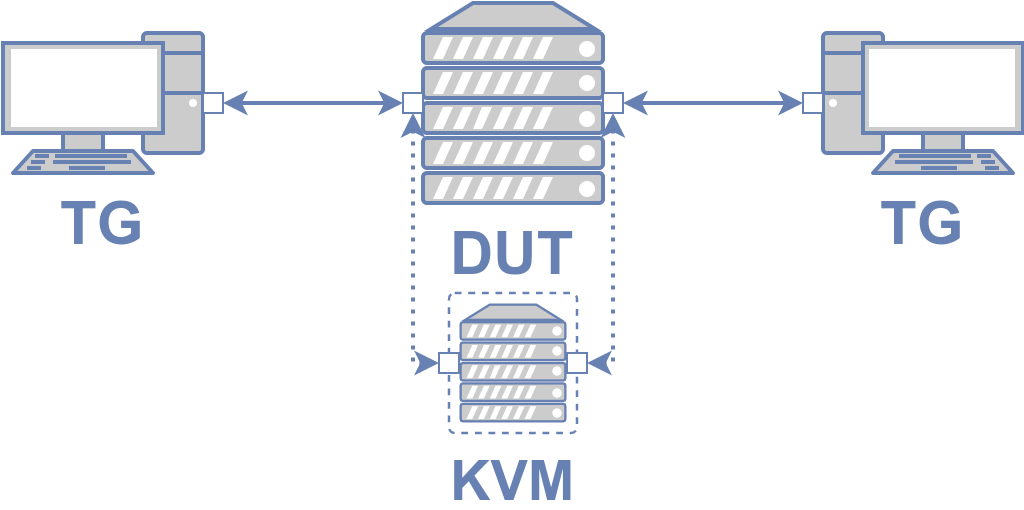}
      \label{fig.performance.setup.bridged}
    }
  \caption{Measurement Setups. TG = Traffic Generator. DUT = Device Under Test.
         Plain arrows are physical connections,
   Dotted arrows are bridge networks and the machine surrounded by dots
   is a virtual machine.}
  \label{fig.performance.setup}
\end{figure}

We configured our testbed into four different setups: A \emph{direct}
client-to-server communication setup, shown in
Fig.~\ref{fig.performance.setup.direct}, that is used to evaluate bandwidth
baselines and rule out sender-bounded experiments. An \emph{indirect} setup,
Fig.~\ref{fig.performance.setup.indirect}, in which the DUT forwards traffic
between sender and receiver by running code on its host OS. A \emph{PCI
passthrough} setup, Fig.~\ref{fig.performance.setup.passthru}, allowing the
hypervisor to directly connect a PCI device, in our case the NIC, with a guest
OS. Finally, a \emph{bridged} setup, Fig.~\ref{fig.performance.setup.bridged},
where the guest OS interfaces are connected to the host OS interfaces using two
bridges.

PCI passthrough relies on hardware virtualization, which is made available by
Intel Virtualization Technology for Directed I/O (Intel VT-d), and on I/O
translation, which is provided by Input-Output Memory Management Unit (IOMMU),
in order to allow guest virtual machines to directly use PCI devices.
We also evaluated the virtio KVM I/O virtualization driver, which allows DPDK to
have fast virtual access to PCI devices, and although more difficult to
configure properly, virtio does not show significant differences in performance.
We did not evaluate Single Root-I/O Virtualization (SR-IOV), but we expect it to
performs similarly.

As mentioned above, we generate traffic using both iperf and wrk with nginx.
Given that iperf relies on a single TCP connection, it is bounded to a single
CPU, and we have to make sure that we measure the performance of the DUT and not
the TGs. To this end, we run a single pair of iperf client-server using the
direct setup, and we add iperf client-server pairs until the bandwidth reaches
the maximum capacity. We found that at least 3 iperf client-server pairs are
needed to allow us to reach a consistent 37.7~Gbps bandwidth, which is the
closest that iperf can get to the maximum capacity of the NICs. We arbitrarily
choose to use 7 iperf pairs for the rest of the experiments. This experiment
aims at analyzing the effect of a small amount of large flows, whose processing
cannot be distributed on all available DUT CPUs.

The wrk+nginx traffic generators consists in one TG running nginx, hosting files
of different size (1KB, 2KB, 4KB, 8KB, 16KB, 32KB, 64KB, 128KB, and 256KB), and
the other TG running 32 wrk threads, each opening 128 connections and transferring the same file. We notice that, when transferring files from 1~KB to 32~KB, the bandwidth is linearly increasing. Starting from 32-KB files, the bandwidth reaches a threshold that holds for experiments transferring files from 64~KB to 256~KB. We arbitrarily choose to transfer the 128KB file for all wrk+nginx experiments. At line-rate, wrk generates an average of 35,780 requests/sec.

For both iperf and wrk+nginx traffic generation, the experiments last for 20 seconds and omit the first second, to avoid transient effects. Packets are sized according to Ethernet MTU. All NICs distributes packets to the RX rings by hashing both IP addresses and ports. Each experiment result is averaged over ten runs for bandwidth measurements, and a thousand runs for RTT and CPU measurements.

\subsection{Experiments}\label{experiments}
The experiments consist in comparing \mmb to trending high-speed packet
processors: FastClick~\cite{barbette2015fast}, XDP~\cite{hoiland2018express},
and iptables~\cite{iptables}. We evaluate two Linux kernel versions (i.e., 4.9
and 4.15) for both \mmb and iptables because they exhibit significant
performance differences. Below, we describe the compared tools.

FastClick~\cite{barbette2015fast} is a packet processor framework based on the
Click modular router~\cite{kohler2000click}. It comes with multi-queue support,
zero-copy forwarding, I/O and computation batching, and integrates both
DPDK~\cite{dpdk-link} and Netmap~\cite{rizzo2012netmap}.It emerged from the
analysis and integration of the best ideas of previous work such as
RouteBricks~\cite{dobrescu2009routebricks} and DoubleClick~\cite{kim2012power},
plus a few additional performance improvements. It also greatly eases the writing of Click configurations, as the framework can handle some level of parallelization automatically, without requiring the user to allocate resources manually as in the other Click-based frameworks.

eXpress Data Path (XDP)~\cite{hoiland2018express} is a high-performance
programmable kernel packet processor for Linux. It does not replace the
TCP-IP stack, but rather adds an extra filtering step based on extended
Berkeley Packet Filters (eBPF). The latters are able to perform stateless
lookups, flow lookups, and flow state tracking. The main use cases of XDP
are pre-stack DDoS filtering, forwarding, load balancing, and flow monitoring.
There is a proposal for the migration of Linux iptables to eBPF/XDP-based
filtering~\cite{karlsson2018path,brouer2018xdp}. We point out that it is
possible to use XDP alongside VPP and \mmb. Because eBPFs are introduced
in the 4.14 kernel, we only evaluated XDP on the Ubuntu Server 18.04 DUT.

iptables is the builtin Linux firewall application~\cite{iptables}. It relies on
netfilter, the kernel packet filtering framework, which consists in multiple
filtering hooks (\texttt{NF\_IP\_LOCAL\_IN}, \texttt{NF\_IP\_FORWARD},
\texttt{NF\_IP\_LOCAL\_OUT}) positioned strategically in the networking stack,
that are triggered by packets as they progress in the stack. However, the
filtering is performed sequentially and the packets that matches drop rules are
not necessarily dropped immediately and might stay longer in the processing
pipe. iptables also comes with a connection tracking system
(\texttt{conntrack}), implemented on top of netfilter.

The following use cases are considered for comparing the performances of \mmb to
FastClick, XDP, and iptables: packet forwarding, firewall-like packet filtering,
packet filtering with stateful flow tracking, packet filtering with stateful
matching and packet mangling (i.e., NAT), and TCP options filtering and
mangling. We also report additional results showing the limitations of \mmb
matching algorithm, and an analysis of its CPU time.

\subsection{Results}\label{results}

\subsubsection{Forwarding}\label{performance:forwarding}

\begin{figure}
\centering
  \includegraphics[width=8.5cm]{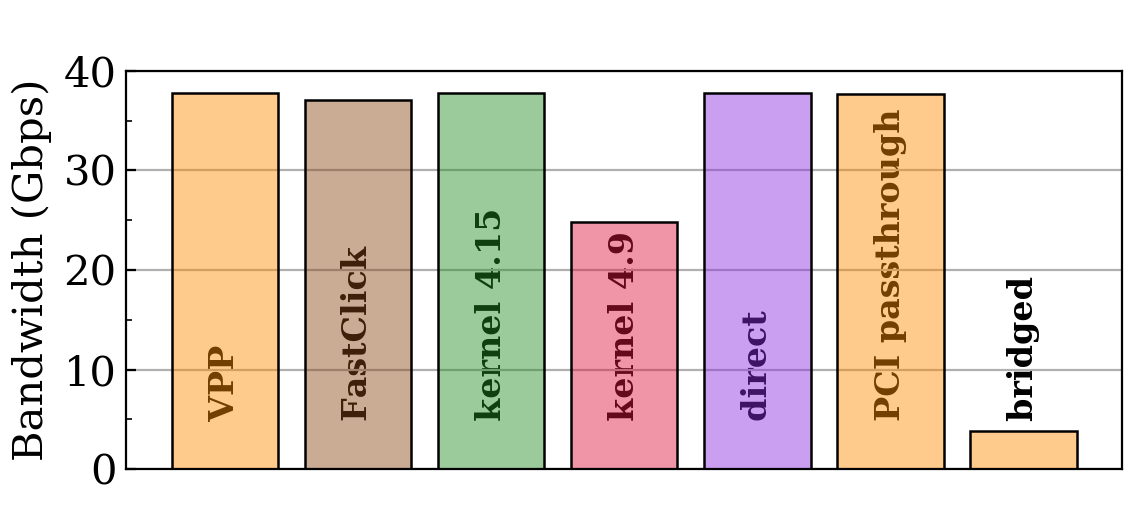}
  \caption{Forwarding baselines. TGs are running iperf. In indirect setup, DUT
  is forwarding packets through either VPP, FastClick, kernel 4.9 or 4.15.
  In PCI passthrough and bridged setups, DUT is running VPP.}
  \label{fig.performance.baseline}
\end{figure}

We first evaluate the TG bottleneck, by running both experiments using the
\emph{direct} setup. We obtain 37.7~Gbps with both iperf and wrk with nginx.
Then, we evaluate VPP, FastClick, and kernel forwarding baselines for the
\emph{indirect} setup, and the VPP forwarding baseline for the
\emph{PCI passthrough} and \emph{bridged} setups. The results are displayed in
Fig.~\ref{fig.performance.baseline}.

We observe that VPP, FastClick, and Linux kernel 4.15 forward packets at more
than 99\% of the direct baseline. The Linux kernel 4.9 performs substantially
worse, forwarding only at 24.8~Gbps. We conducted additional analyses and ruled out unfortunate queue balancing, CPU loads, and RX input hash methods as the causes of this difference.

When running VPP, both the \emph{indirect} and \emph{PCI passthrough} setups
reach the \emph{direct} baseline. Both setups continue to behave similarly in
following experiments. We note that this advocates in favor of \mmb deployment
flexibility and from now on, we report a single result that stands for both
setups. Unsurprisingly, the \emph{bridged} performs very poorly at 3.6~Gbps,
emphasizing so the importance of direct I/O.

\begin{figure}
  \begin{center}
   \hspace*{-0.5cm}
    \includegraphics[width=7.5cm]{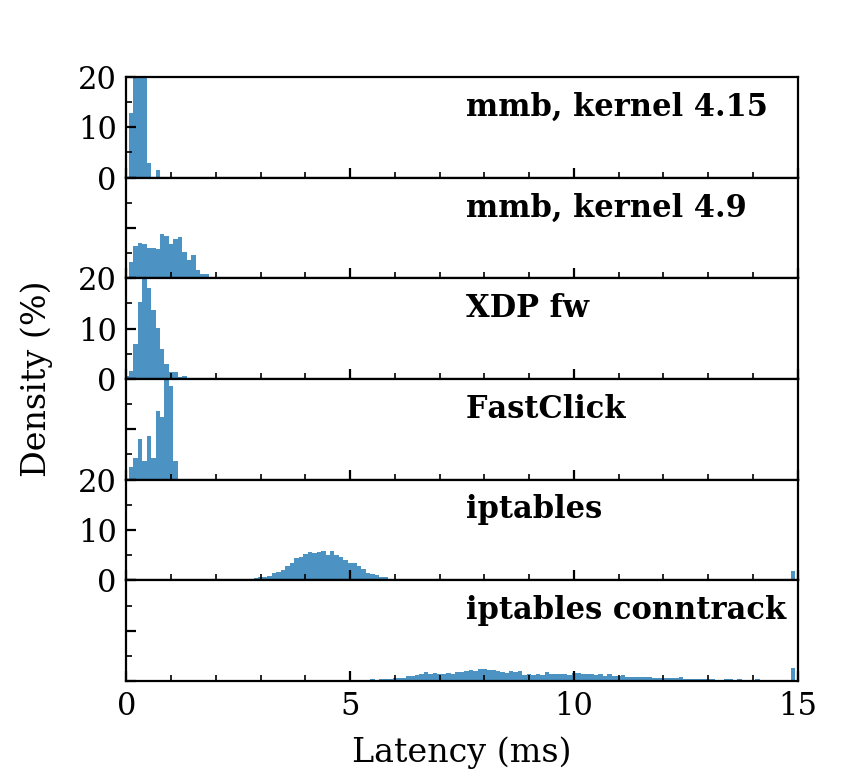}
  \end{center}

  \caption{RTT with 10K firewall rules.}
  \label{fig.firewall.rtt}
\end{figure}

\begin{figure*}[htp]

	\captionsetup{type=figure}\addtocounter{figure}{+1}

\makebox[\textwidth][c]{%
    \begin{minipage}{0.43\textwidth}
      \centering
      \includegraphics[width=7.0cm]{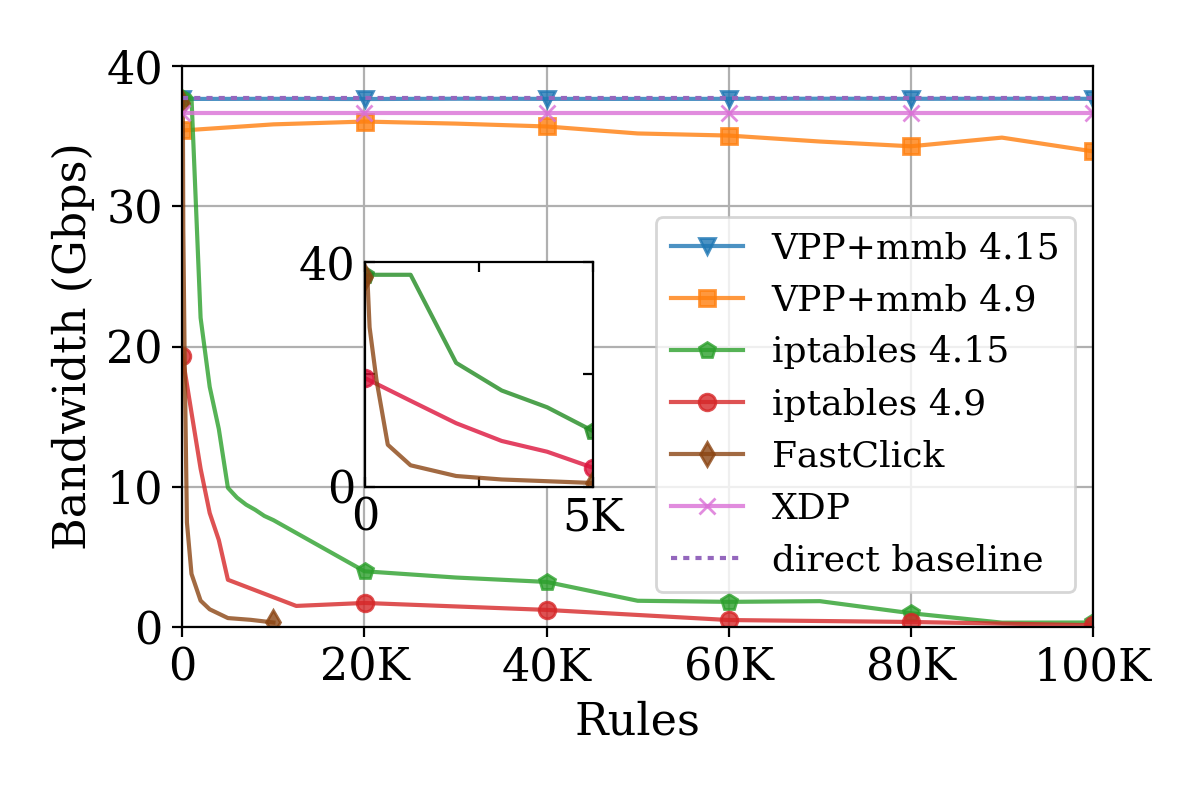}
      \vspace*{-0.5cm}
      \captionof{subfigure}{Firewall 5-tuple filtering, iperf}
      \label{fig.performance.baremetal.firewall.iperf}
    \end{minipage}%
   \hspace*{-1.0cm}
    \begin{minipage}{0.43\textwidth}
      \centering
      \includegraphics[width=7.0cm]{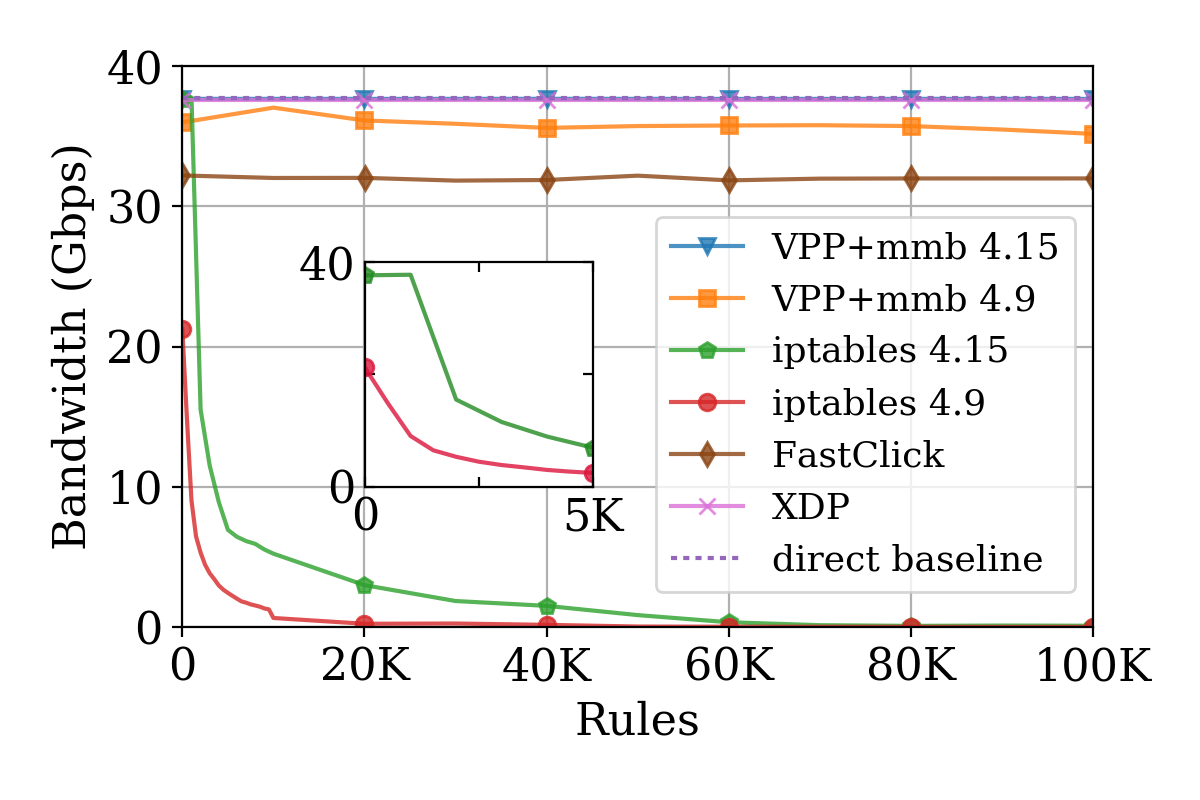}
      \vspace*{-0.5cm}
      \captionof{subfigure}{Stateful matching, iperf}
      \label{fig.performance.baremetal.stateful.iperf}
    \end{minipage}%
   \hspace*{-1.0cm}
    \begin{minipage}{0.43\textwidth}
      \centering
      \vspace*{-0.25cm}
      \includegraphics[width=7.0cm]{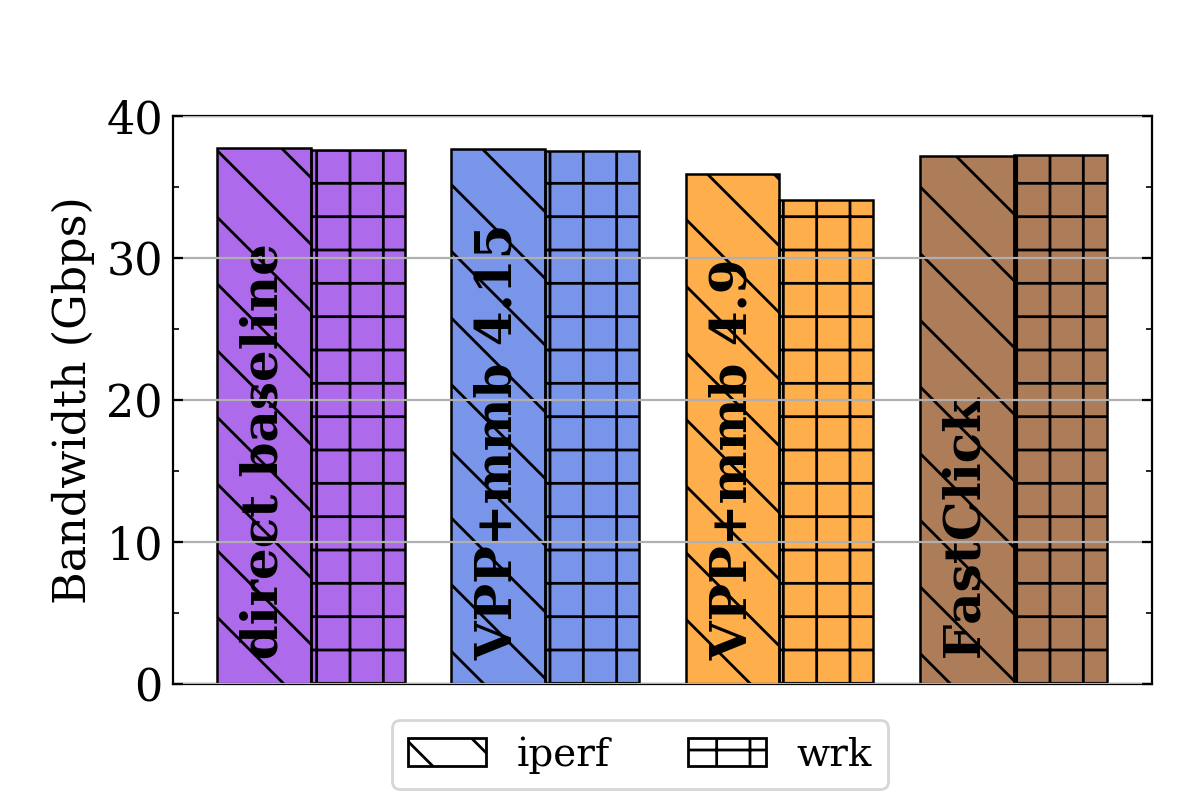}
      \vspace*{-0.1cm}
      \captionof{subfigure}{NAT}
      \label{fig.performance.baremetal.nat}
    \end{minipage}%
  }%

  \makebox[\textwidth][c]{%
    \begin{minipage}{0.43\textwidth}
      \centering
      \includegraphics[width=7.0cm]{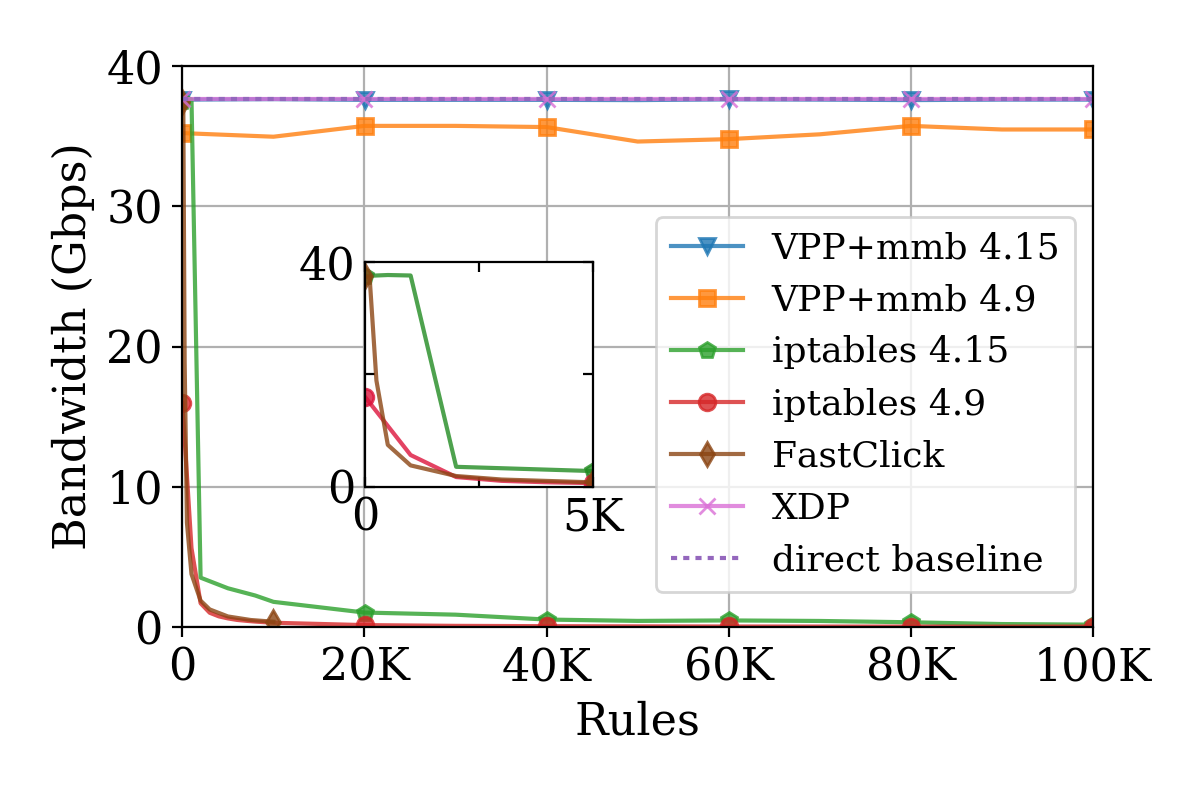}
      \vspace*{-0.5cm}
      \captionof{subfigure}{Firewall 5-tuple filtering, wrk+nginx}
      \label{fig.performance.baremetal.firewall.wrk}
    \end{minipage}%
    \hspace*{-1.0cm}
    \begin{minipage}{0.43\textwidth}
      \centering
      \includegraphics[width=7.0cm]{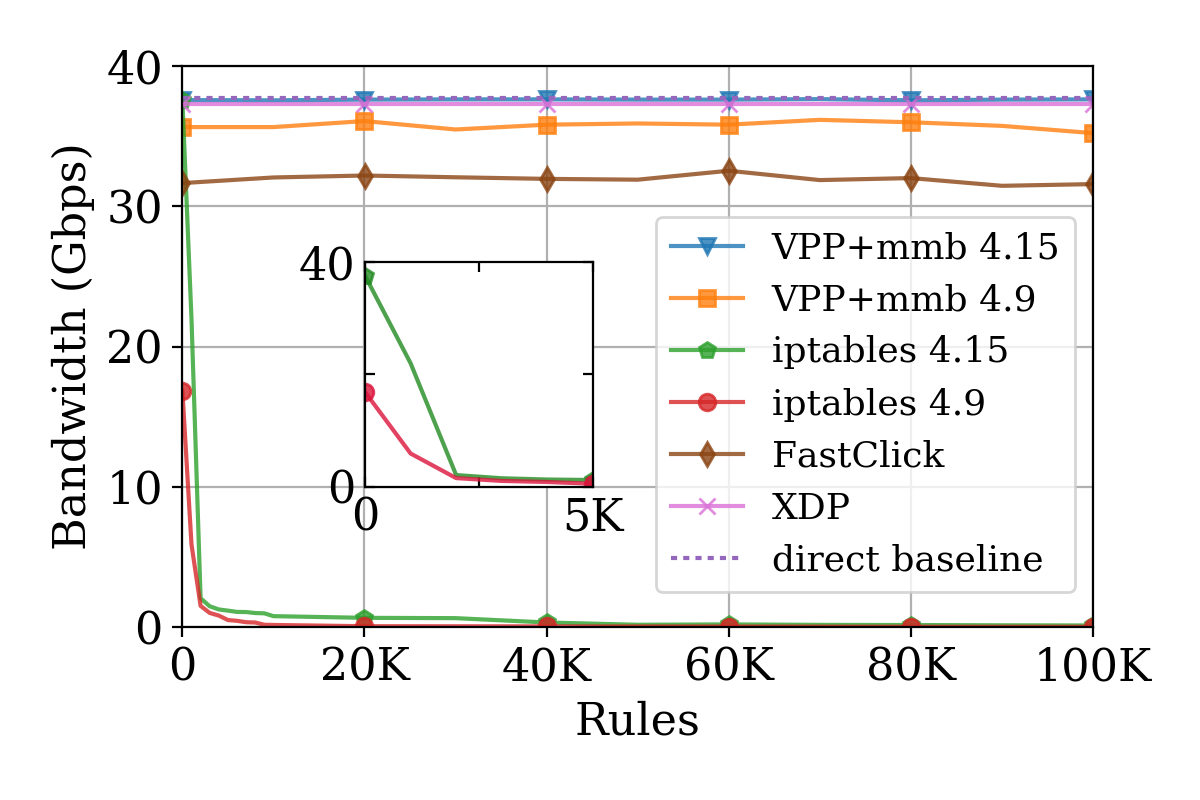}
      \vspace*{-0.5cm}
      \captionof{subfigure}{Stateful matching, wrk+nginx}
      \label{fig.performance.baremetal.stateful.wrk}
    \end{minipage}%
    \hspace*{-1.0cm}
    \begin{minipage}{0.43\textwidth}
      \centering
      \includegraphics[width=7.0cm]{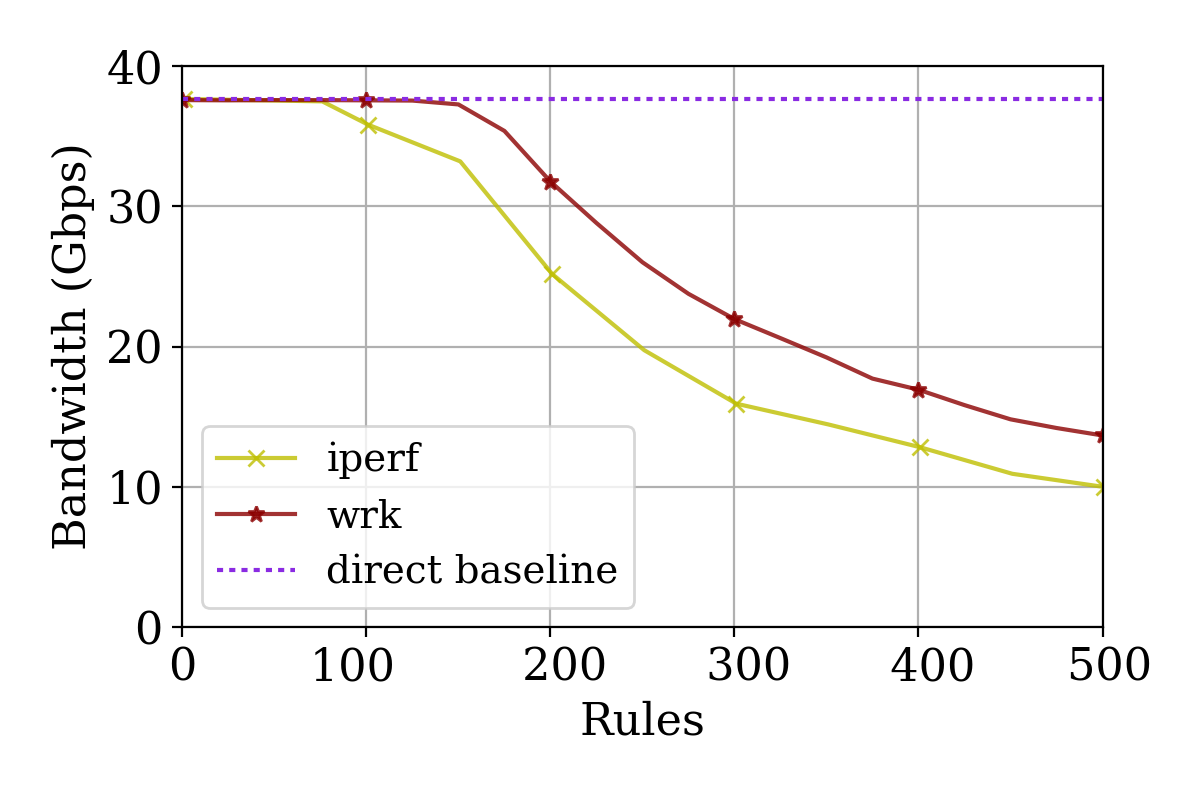}
      \vspace*{-0.5cm}
      \captionof{subfigure}{TCP Options matching.
                VPP+\mmb on a DUT\\ running a Linux kernel version 4.15.}
      \label{fig.performance.baremetal.opt}
    \end{minipage}%
  }%

  \captionsetup{type=figure}\addtocounter{figure}{-1}

  \caption{Performances in \emph{indirect} setup.
  \emph{direct} baseline traffic does not traverse the DUT.}
  \label{fig.performance.bw}
\end{figure*}

\subsubsection{Firewall}\label{performance:firewall}
We configure \mmb as a firewall and compare it to a FastClick firewall
configuration, an XDP-based firewall, and the kernel forwarding with iptables
filtering, to evaluate their applicability to a basic firewall-like packet
filtering use case. To this end, a generator of random firewall rules is used.
Rules classify packets based exclusively on five-tuples, to enable tools with
mask-based hash classification approaches (e.g., \mmb and XDP) to only use one
single table. We ensure that no rules are matching the traffic from the TGs. We
generate a new set of rules for every experiment. In the real world, these
middlebox policies can be used as a firewall as well as DDoS protection
measures.  We inject these rules to \mmb as stateless rules.

For XDP, we build a simplistic five-tuple firewall. We use a BPF hash map with
five-tuples as keys, and use it to store the type of rule (i.e., accept or drop)
and the count of accepted and dropped packets. Every time a packet is received,
an eBPF will check for an entry in the map corresponding to the current packet.
If it is found, then the related drop counter is incremented and the packet
dropped. Otherwise, the packet passes.

When the DUT is running iptables, we inject the rules to the \texttt{FORWARD}
chain.

For FastClick, we use an \texttt{IPFilter} element that will drop packets
matching a rule (i.e., none in our test), and will pass them to the routing
table otherwise. As \texttt{IPFilter} elements can only support up to $2^{16}$
rules, we have to chain several of those to support more rules. The
\texttt{click-fastclassifier} post-processing tool was not used. While it
moderately improves performance, it takes a huge amount of time and RAM to
optimize the configuration, and makes it static (i.e., preventing so live
reconfiguration).

The bandwidth results are shown in
Fig.~\ref{fig.performance.baremetal.firewall.iperf} and
\ref{fig.performance.baremetal.firewall.wrk}. It shows that XDP and VPP with
\mmb on a 4.15 kernel, keep a constant forwarding rate, regardless of the number
of rules. Both are performing very close to the direct baseline, at the
exception of XDP on iperf-generated traffic, that has a large standard
deviation (not represented in this figure) which we believe is the effect of
unfortunate CPU distribution. This effect is not present with wrk traffic,
because it generates more flows, whose processing distributes better on
multiqueue systems. The \mmb firewall on a 4.9 kernel shows signs of rule count
dependent performance, but we believe this is rather due to the kernel.

iptables on a 4.15 kernel surprisingly sustains a line-rate bandwidth until
1,000 rules are inputted, while iptables on 4.9 kernel performance decreases
already with very few rules.

FastClick performance decreases even more quickly with the number of rules (no
data is depicted for more than 10,000 rules because the slow processing stalls
the TGs). This is due to the implementation of the \texttt{IPFilter} element
that, as already noted, is not designed for a large number of rules. On the
contrary to \mmb and XDP that use a $\mathcal{O}(1)$ hash-based approach to
match packets to rules, FastClick uses a binary search which requires
$\mathcal{O}(\log_2 n)$ comparisons. Moreover, the matching code has bad cache
locality, further contributing to the performance drop.

This experiment indicates that the \mmb mask-based \emph{fast path}, when
relying on a single table, has a very limited impact on the maximum achievable
bandwidth of the forwarding device, regardless on the number of rules.

In Fig.~\ref{fig.firewall.rtt}, we display the average RTT distribution
for the same use case, with 10,000 rules. It shows that all DPDK-based I/Os,
plus XDP, have an average RTT under 1~ms. At the microsecond scale, \mmb
on a 4.15 kernel performs better with an average RTT of 327~$\mu$s, against
562~$\mu$s for XDP and 771~$\mu$s for FastClick.

\subsubsection{Stateful}\label{performance:stateful}
Next, we confront the chosen tools to the packet filtering with stateful flow
tracking use case. We generate sets of rules matching on the received packets
five-tuple, similarly to the previous experiment. In addition to that, we input
a static set of rules to guarantee that all traffic from the TGs is matched, to
enable flow tracking capabilities of the tested tools, and quantify the inducted
performance overhead. In the real world, this type of middlebox policies can be
used for private network-initiated reflexive ACLs.

We inject these rules to \mmb as stateful rules in order to have every packet
matching at least one rule to add an entry to the connection table. All packets,
matching or not, are also checked against the opened connections, whose states
are updated when needed.

For FastClick, we use an \texttt{IPRewriter} element for connection tracking.
Each packet goes through that element when leaving the middlebox, triggering the
creation of a new flow entry if it is a flow first packet. Packets also go
through that element just after entering the middlebox. If a packet matches an
existing flow, it is passed directly to the routing table rather than to the
\texttt{IPFilter}. \texttt{IPRewriter} is not thread-safe and will only
recognize a return packet as belonging to a flow if it is processed on the same
core that created the flow entry. As Receive-Side-Scaling cannot enforce that,
we use one \texttt{IPRewriter} per core and keep separate flow state for both
directions.

iptables is configured as a stateful firewall by enabling \texttt{conntrack},
the iptables netfilter-based connection tracking module, and injecting rules to
the \texttt{FORWARD} chain.

With XDP, we build a stateful flow tracker using three different BPF hash maps.
One for connections tracking, one for three-tuple matching rules and one for
five-tuple matching rules. We ensure that all operations are as simple as
possible to avoid unnecessary processing overhead. The five-tuple map is filled
with the randomly generated rules, and the three-tuple map with the static
rules. Both maps are used to filter packets and add entries to the tracking map.
The tracking map contains the list of forwarded flows that matched at least one
rule, with hashes of the five-tuples as keys. This map also stores various
pieces flow information such as flow timestamps, TCP state, packet counters, and
involved interfaces. Since a full TCP state machine is superfluous for this use
case, we implement a simplified version by observing all TCP flags sent from
both ends of every connection, which is enough to differentiate transient from
idle connections.

Results are displayed in Fig.~\ref{fig.performance.baremetal.stateful.iperf}
and \ref{fig.performance.baremetal.stateful.wrk}. As in the previous experiment,
\mmb on a 4.15 kernel and XDP have constant line-rate performances. iptables
on a 4.15 also shows similar performance than for the stateless firewall
experiment, while iptables kernel 4.9 with \texttt{conntrack} performs
worse than without it.

FastClick performs better than for the stateless case, because the costly
filtering step is done only for the first packet of each flow (but in both
directions). Its performance is still significantly lower than that of \mmb or
XDP, however.

\subsubsection{NAT}\label{performance:nat}
The NAT experiment consists in DUT running a Source Network Address
Translation (SNAT) with port translation. The straightforward way to configure
\mmb as a SNAT, with 200.0.0.1 as the globally routable address, relies on
the single following rule:
\begin{lstlisting}[style=BashInputStyle]
vpp# mmb add-stateful ip-saddr 10.0.0.0/24 ip-proto tcp tcp-syn shuffle tcp-sport mod ip-saddr 200.0.0.1
\end{lstlisting}

For FastClick, we use \texttt{IPRewriter} elements to implement the SNAT. When
a packet leaves the DUT towards the server, it goes through an
\texttt{IPRewriter} which will rewrite its source address and port, adding a
new flow entry if necessary. The port is chosen in a range that depends on the
processing core so that on the return path, the packets can be dispatched to
the same core for the reverse transformation using a simple match on the
destination port.

As shown in Fig.~\ref{fig.performance.baremetal.nat}, the average bandwidth
of traffic crossing an \mmb NAT is equals to the direct baseline. \mmb performs
1 and 2.5~Gbps worse on a 4.9 kernel, which we believe is directly caused by
a constant kernel overhead, observed on all other use cases. FastClick performance is very similar. While the Click configuration for the NAT also comprises a slow \texttt{IPFilter} element, it is used only on the return path, and contains just one rule per core. The NAT performance is thus better than the
one of the stateful firewall.

\subsubsection{TCP Options}\label{performance:tcpoptions}
Finally, we evaluate the performance of traffic engineering policies that
matches and mangle TCP Options. The processing of TCP Option, or IPv6 Extension
Headers, is more complex because it requires linked list parsing for every packet.
Because any ingress middlebox is able to strip or add TCP Options, the presence
and order of TCP Options in a TCP packet is not known a-priori. This forbids
\mmb to exploit a rule-defined or connection-defined mask-based approach for
parsing TCP options. For this use case, we inject rules that match on random
value of random TCP Options. We do not mangle TCP options because it would
disrupt TCP and affect its performance. However, because the linked list
parsing is done once, we predict that mangling option do not increase
substantially the processing time.

FastClick is not tested against this use case because none of the distributed
elements is able to match on variable-offset TCP options. We would have to
write new elements for TCP option parsing and mangling, which would require a
significant amount of code.

XDP is also not tested against this use case because eBPF brings limitations
explained in the next section, which are exceeded by the task of implementing
complex packet mangling.

The bandwidth measurement results,
displayed in Fig.~\ref{fig.performance.baremetal.opt}, indicates that
the threshold of injected TCP Options-based classification rules to sustain
line-rate packet forwarding, for traffic generated with iperf, and wrk with
nginx, are respectively 100 and 150. This difference is explained by the
high CPU-time requirements of this use case, that we investigate in
Sec.~\ref{performance:cpu}, which makes it benefits from multi-core processing.

\subsubsection{Limitations}\label{performance:limitations}

\begin{figure}
    \centering
   \includegraphics[width=7.0cm]{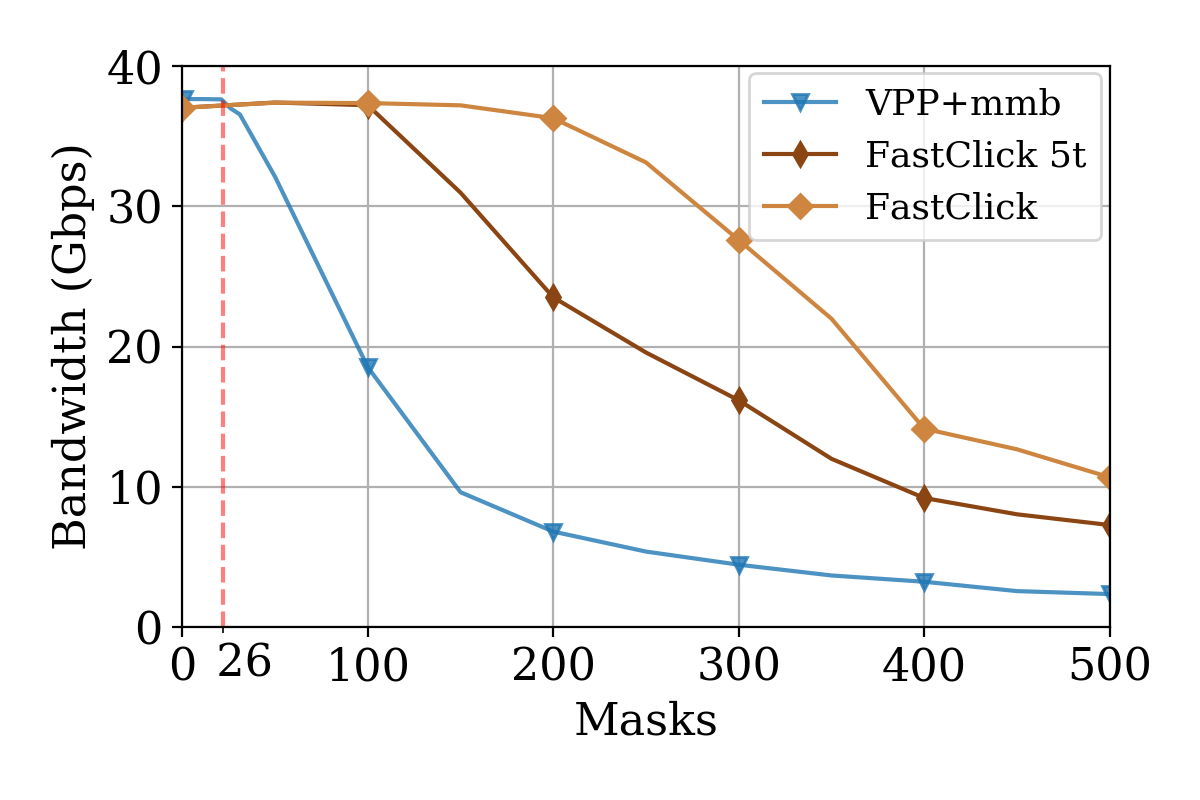}

   \label{fig.performance.baremetal.masks}
  \caption{Stability limit of \mmb mask-based matching.
           FastClick 5t is the firewall matching on 5-tuples.
           FastClick is the firewall matching on random rules.}
  \label{fig.performance.baremetal}
\end{figure}

We conducted an additional experiment on the firewall use case to quantify
the performance gain of \mmb mask-based approach
(see~\ref{architecture:classification}). In this scenario, each inputted rule
is deliberately generated to match on a different combination of five fields, in
order to force the use of one table for each single rule. The results are shown
in Fig.~\ref{fig.performance.baremetal}.

It shows a clear limit of 26 combinations of fields for line-rate processing,
after which the performances start to diminish. With more than 40 combinations,
performances drop significantly faster. We explain it by the limited size of the
cache that is too small to hold all prefetched hash tables to avoid cache miss.
We advocate that this limitation is largely sufficient for a realistic usage.

XDP runs almost at line rate for both firewall
(Fig.~\ref{fig.performance.baremetal.firewall.iperf} and
Fig.~\ref{fig.performance.baremetal.firewall.wrk}) and stateful
(Fig.~\ref{fig.performance.baremetal.stateful.iperf} and
Fig.~\ref{fig.performance.baremetal.stateful.wrk}) use cases, which makes it a
good alternative to \mmb. But, although very good, it introduces some
limitations and restrictions. Indeed, the in-kernel eBPF verifier is a static
code analyzer that walks BPF programs instruction per instruction, and validates
them. Moreover, stack space in BPF programs is limited to only 512 bytes which
means that any program must terminate quickly and will only call a fixed number of kernel functions. As a consequence, larger programs or programs that contain
loops will be rejected, which provides security and reliability for the kernel
but can be a drag on developers.

While FastClick does not support large numbers of rules, we observe that its
matching algorithm, based on a decision tree, is not affected much by the number
of combinations of fields it is matching on, and it can sustain 35~Gbps up to 200 rules with different masks. The performance is even slightly better than the one for the five-tuple firewall, since there is more variety in the rules. E.g., if a packet is a TCP packet, all rules for other protocols (ICMP, UDP, etc.) can be ruled out with a single comparison of the IP protocol field.

\subsubsection{CPU time}\label{performance:cpu}
\begin{figure}
  \begin{center}
   \hspace*{-1.0cm}
    \includegraphics[width=8.5cm]{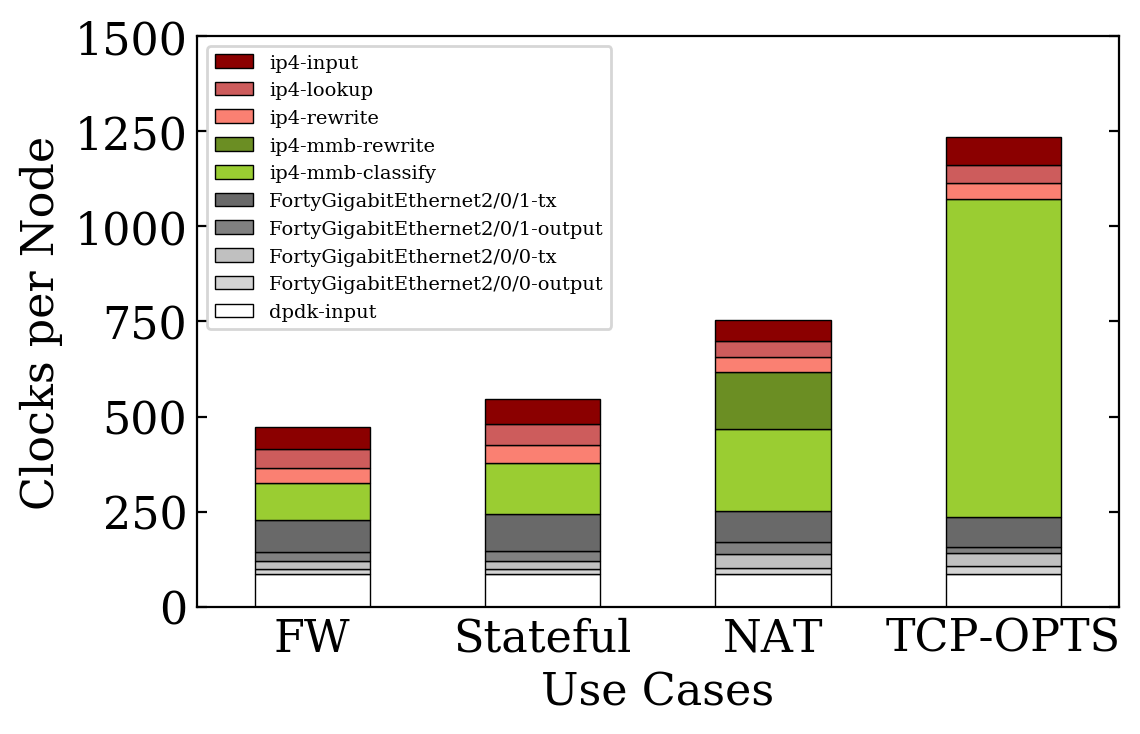}
  \end{center}

  \caption{CPU clock cycles per packet. fw is a 5-tuples firewall with 100K rules, stateful is a flow tracker with 100K rules and one guaranteed match,
  NAT is a source NAT, tcp-opts is a TCP option classifier and rewriter
  with 100 rules.}
  \label{fig.runtime}
\end{figure}

We measured the cost in CPU cycles of packet processing, which is available
through VPP API, for four extreme use cases: respectively a firewall with 100K
rules, a flow tracker with a 100K rules and a guaranteed match, a source NAT and
a transport-level engineering middlebox that parses and modify TCP options that
contains 100 rules. The result, presented in Fig.~\ref{fig.runtime},  shows
that, for the first three experiments, performing combinations of packet
filtering, stateful tracking, and packet mangling rules, \mmb is able to
maintain a low rate of CPU cycles per packet. In particular, packet filtering and flow tracking consume less CPU time, respectively 96 and 135, than the rest of VPP IP processing path (\texttt{ip4-input},\texttt{ip4-lookup}, and
\texttt{ip4-rewrite}), whose execution requires an average of 155 CPU cycles
per packet. Finally, packet classification based on TCP Options consumes the most CPU time, i.e., 836 clocks per packet. We remind that this experiment is an
extreme use case that corresponds to the most complex task that can be performed by \mmb while sustaining line-rate forwarding. Moreover, realistic use cases are not likely to require as many different policies.

\section{Related Work}\label{related}
Over the years, numerous works have been proposed for fast and efficient packet
processing.  Among others, one can cite \dfn{iptables}~\cite{iptables} (the
built-in Linux firewall application), \dfn{PF\_RING}~\cite{deri2004improving} (a
software I/O framework that modifies the socket API to avoid buffer reallocation
and bypass unnecessary kernel functionalities to improve the performances of
packet capture from those of libpcap),
\dfn{PacketShader}~\cite{han2011packetshader} (a GPU-accelerated software router
framework, that perform I/O batching and kernel bypass) and
\dfn{eXpress Data Path} (XDP)~\cite{hoiland2018express}, a
high-performance programmable kernel packet processor for Linux.

The more specific \dfn{mOS}~\cite{jamshed2017mos} is a networking stack for
building stateful middleboxes.  Its ambition is to provide a high-performance
general-purpose flow management mechanism.  It comes with an API to allow for
building middleboxes applications requiring flow state tracking such as stateful
NATs, or payload reassembly such as NIDS/NIPS and L7 protocol analyzers. mOS is
based on mTCP~\cite{jeong2014mtcp}, a parallelizable userspace TCP/IP stack.

The \dfn{Click} modular router is a flexible router
framework~\cite{kohler2000click}.  It was not specifically designed for
high-speed packet processing as it relies on the Linux kernel via system calls
for certain tasks, leading so into an increase in processing time.  Further, on
the contrary to \mmb, Click requires the user to write C++ classes to build new
functionalities.  Click has been extended over the years to overcome its
limitations. \dfn{RouteBricks}~\cite{dobrescu2009routebricks} brings hardware
multiqueue support to Click, and introduce an architecture for parallel
execution of router functionalities as a first step towards fast modular
software routers. \dfn{DoubleClick}~\cite{kim2012power} integrates PacketShader
I/O batching and computation batching. Moreover, it also takes advantage of the
non-uniform memory access (NUMA) CPU architecture.
\dfn{FastClick}~\cite{barbette2015fast} comes with multi-queue support,
zero-copy forwarding, I/O and computation batching, and integrates both
DPDK~\cite{dpdk-link} (user-level packet I/O framework) and
Netmap~\cite{rizzo2012netmap} (one of the earliest userlevel packet I/O
framework). FastClick comes with a wide variety of elements, and is able to
implement more diverse policies than \mmb. Moreover, one of FastClick key point
is the ease of writing new Click configurations. However, the process of
choosing and building a suited element pipeline still require substantial
effort and expertise. \dfn{MiddleClick}~\cite{barbette2018building} further
enhances FastClick with flow-processing capabilities.  It comes, among others,
with an optional middlebox-oriented TCP stack. Moreover, MiddleClick speeds up
NFV chains by factorizing tasks such as classification, which is done only once
for a whole service chain, and can be offloaded to hardware.  Finally,
\dfn{ClickNP}~\cite{li2016clicknp} is an FPGA-based packet processing framework
whose abstraction mimics Click.  ClickNP is supposed to improve parallelism, and
shows extreme performance improvement with more than 200~MPps, but it requires
dedicated expensive hardware.

This paper has shown that VPP with \mmb performs better than those state of the
art solutions for fast packet middlebox processing.

\section{Conclusion}\label{conclusion}
For now ten years, we observe that middleboxes have become more and more popular
at every floor of the network (corporate, mobile networks, tier-1 ASes).  Those
middleboxes have to deal, also, with an increasing Internet traffic, meaning that they have to process packets at very high rate.

This paper introduced \mmb (Modular Middlebox), a high-performance modular
middlebox, implemented as a VPP plugin. \mmb can be used to deploy out-of-the
box middleboxes and to easily and intuitively configure custom policies through
its command-line interface, on the contrary to state-of-the-art solutions usually requiring dedicated hardware, specific OS or non-trivial programming.

We compared \mmb to other trending high-speed packet processors (FastClick,
XDP, iptables) and demonstrated, through several use cases, that \mmb is able
to sustain packet forwarding at line-rate speed when applying a large number of
diverse and complex classification and mangling rules.  \mmb is open source and
freely available.\footnotemark[1]

In the near future, we would like to push further the \mmb development.  In
particular, we are interested in Layer-7 payload reassembly and generic
application-level matching and mangling rules.

\section*{Acknowledgments}
{\small
This project has received funding from the European Union's Horizon 2020
research and innovation program under grant agreement No 688421.
The opinions expressed and arguments employed reflect only the authors' views. The European Commission is not responsible for any use that may be made of that information. Further, the opinions expressed and arguments employed herein do not necessarily reflect the official views of the Swiss Government.
}

\balance

\end{document}